\documentclass[conference]{IEEEtran}
\usepackage{amsfonts}
\usepackage[dvips]{graphicx}
\usepackage{epsfig,latexsym}
\usepackage{float}
\usepackage{indentfirst}
\usepackage{amsmath}
\usepackage{amssymb}
\usepackage{times}
\usepackage{subfigure}
\usepackage{stfloats,algorithm,algorithmic,bm}
\usepackage{psfrag}
\usepackage{cite}
\usepackage{subfigure}
\usepackage{stfloats}
\usepackage{bm}
\usepackage{colortbl}
\usepackage{xcolor}
\usepackage[section]{placeins}

\begin{document}

\title{Joint Multi-Cell Resource Allocation Using Pure Binary-Integer Programming for LTE Uplink}

\author{\IEEEauthorblockN{Tong Zhang, Xiaofeng Tao,  Qimei Cui}
\authorblockA {Key Laboratory of Universal Wireless Communication, Ministry of Education}
Beijing University of Posts and Telecommunications, Beijing, 100876, China\\
Email: bennytongzhang@gmail.com}
\maketitle

\maketitle

\begin{abstract}
\boldmath
Due to high system capacity requirement, 3GPP Long Term Evolution (LTE) is likely to adopt frequency reuse factor 1 at the cost of suffering severe inter-cell interference (ICI). One of combating ICI strategies is network cooperation of resource allocation (RA). For LTE uplink RA, requiring all the subcarriers to be allocated adjacently complicates the RA problem greatly.
This paper investigates the joint multi-cell RA problem for LTE uplink. We model the uplink RA and ICI mitigation problem using pure binary-integer programming (BIP), with integrative consideration of all users' channel state information (CSI). The advantage of the pure BIP model is that it can be solved by branch-and-bound search (BBS) algorithm or other BIP solving algorithms, rather than resorting to exhaustive search. The system-level simulation results show that it yields 14.83\% and 22.13\% gains over single-cell optimal RA in average spectrum efficiency and 5th percentile of user throughput, respectively.
\end{abstract}

\section{Introduction}
\IEEEPARstart{L}{o}calized single-carrier frequency division multiple access (SC-FDMA) is applied to 3GPP Long Term Evolution (LTE) uplink system requires only adjacent subcarriers can be allocated to users. One of advantage is mitigating the peak-to-average-power ratio problems and user equipment implementation complexity \cite{b1}. However, subcarrier adjacency constraint makes resource allocation (RA) problem much more complex than orthogonal frequency division multiple access (OFDMA) system \cite{b2}.

The inter-cell interference (ICI) is one of the main interference sources in the LTE uplink, especially degrading cell-edge users' performance. The partial frequency reuse (PFR) and soft frequency reuse (SFR) (frequency reuse greater than 1) will alleviate ICI, but reduces average spectrum efficiency \cite{b3}. In order to achieve high system capacity, the LTE uplink system will probably adopt frequency reuse 1 strategy \cite{b4}. In this situation, one effective inter-cell interference coordination (ICIC) method is that base stations (BSs) jointly allocate resources, with the help of a centralized scheduler and high-capacity backhaul. The optimal case is joint power control (PC) and RA, but greatly increases the computational complexity. \cite{b5} solves the problem for OFDMA system using the mixed-integer programming (MIP) by exhaustive search. We consider one way of lowering complexity is splitting the problem into PC and RA \cite{b6}. Furthermore, the open-loop PC is taken \cite{b6}, which is an inverse PC for combating near-far effect and similar to LTE protocols \cite{b7}.

The RA and PC strategies are investigated in device-to-device \cite{b8} and relay selection \cite{b9}, respectively. A multi-cell cooperative RA scheme for LTE uplink using interference-aware strategy based on proportional fair (PF) is proposed in \cite{b6}. In \cite{b6}, a centralized scheduler not only measures the interference current cell suffered but also evaluates influence of forthcoming scheduling decisions to scheduled cells. However, that scheme does not take all channel state information (CSI) into integrative consideration. The average spectrum efficiency and cell-edge user throughput can continue to boost, if joint multi-cell RA is carried out in a more comprehensive CSI utilization.

In this paper, we introduce a joint multi-cell RA scheme using pure binary-integer programming (BIP). The proposed scheme generates allocation strategy for all cooperative cells simultaneously using comprehensive CSI knowledge.  We aim at maximizing the sum of metric, and create constraints by incorporating the inter-cell user pairs (UPs) with resource patterns. We also analyze the performance of our proposed scheme, multi-cell cooperative RA \cite{b6} and single-cell optimal RA \cite{b11} by means of extensive system-level simulation. Tracing back to \cite{b11}, based on the set partition method, it proposes a kind of pure BIP model to optimize single-cell LTE uplink RA. The merit of pure BIP model is that it can be solved by BBS algorithm or other BIP solving algorithms, without resorting to exhaustive search \cite{b11}. Moreover, \cite{b12} applies it to virtual-MIMO systems, which focuses on how to pair intra-cell users. Through the set partition method, \cite{b12} obtains the optimal solution of virtual-MIMO systems costing lower complexity than the exhaustive search model.  Adding into distributed ICI mitigation, \cite{b1} extends it to multi-cell scenes, which not only generates the single-cell UPs scheduling strategies simultaneously but also mitigates the ICI by exchanging of high interference indicators. For introducing inter-cell UPs alignment for resource constraint, we can apply the pure BIP model to joint multi-cell RA  as a suboptimal scheme. Besides, the joint multi-cell RA is characterized by considering all CSI and turning out all allocation results at the same time. Last but not least, our contributions can be summarized as follows, to the best of our knowledge, it is the first time that the pure BIP model is applied to multi-cell scenes with comprehensive consideration of all CSI.

The remainder of the paper is organized as follows. Section II introduces the cooperative scheduling system for LTE uplink at first, then formulates the optimal joint multi-cell RA model. The pure BIP model is introduced with defining the novel metric for throughput and fairness. Simulation results are shown in Section III and conclusion is drawn in Section IV.
\begin{figure}
\centering
\includegraphics[width=2.2in]{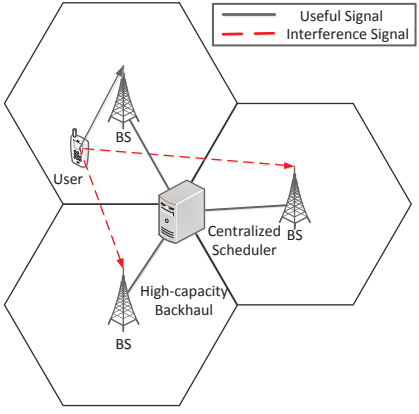}
\caption{Take three cells cooperative scheduling network for example.}
\end{figure}
\vspace{-0.3em}
\section{System Model}
Consider an LTE uplink system with $U$ BSs / cells. The ${uth}$ cell includes ${K_u}$ users and  $F$ resource blocks (RBs). One RB consists of twelve consecutive subcarriers in the frequency domain and one slot duration (0.5msec) in time domain. Each RB only can be exclusive to one user per cell, namely intra-cell pairing is not allowed. Each BS and users are equipped with ${N_r}$ receive antennas and ${N_t}$ transmission antennas respectively. One centralized scheduler knowing all CSI connects with $M$ cells via the high-capacity backhaul. In downlink coordinated multi-point transmission/reception (CoMP), the architecture that the coordinated scheduling/beamforming (CS/CB) transmission scheme also can be taken \cite{b10}.

We use the non-cooperative receiving, without utilization of other cells' leakage signal. A fast infrastructure is employed, where the backhaul capacity can be seemed as large enough. For each cell, we adapt linear minimum mean square error (LMMSE) receiver to equalize the received signal. For practical consideration, the LMMSE receiver enjoys a much lower complexity than maximum likelihood (ML) receiver, but keeping appropriate performance.
Denote ${h_{i,j,q}}$  as the ${ith}$  user channel matrix on the  ${qth}$ subcarrier with the ${jth}$ RB. The covariance matrix of interference plus noise can be expressed by ${R_{zz}}$ \cite{b6}.
\begin{equation}
{R_{zz}} = {\sum\limits_{i' \in {\psi _i}} {{P_{i'}}{{{h_{i',j,q}}{h_{i',j,q}^H}}}} }+ {\rm E}\left[ {{n_{i,j,q}}n_{_{i,j,q}}^H } \right]
\end{equation}
 ${\left(  \cdot  \right)^H }$ is the conjugate-transpose operator. ${n_{i,j,q}}$ is the thermal noise.

Let ${P_i}$ represents transmission power per subcarrier of the $ith$ user. ${\psi _i}$ denotes the set of users scheduled on the $jth$ RB, removing the $ith$ user. Consequently, the LMMSE decoding matrix ${w_{i,j,q}}$ can be written as \cite{b6}:
\begin{equation}
{w_{i,j,q}} = h_{i,j,q}^H \left( {{h_{i,j,q}}h_{i,j,q}^H  + diag{{({R_{zz}})}}} \right)^{ - 1}
\end{equation}
In order to calculate the instantaneous throughput of ${ith}$  user on the $jth$ RB, we should firstly present the expression of the signal-to-noise-ratio (SINR) \cite{b6}.
\begin{equation}
{\gamma _{i,j,q}} = \frac{{{P_i}{w_{i,j,q}}{h_{i,j,q}}h_{i,j,q}^Hw_{i,j,q}^H}}{{{w_{i,j,q}}{R_{zz}}w_{i,j,q}^H}}
\end{equation}
Giving the $ith$ user' SINR on every subcarrier, the instantaneous throughput of $ith$ user on the $jth$ RB can be optimistically estimated via the Shannon formula.
\begin{equation}
{R_{i,j}} = B  \cdot \sum\limits_{q = 1}^Q {{{\log }_2}\left( {1 + {\gamma _{i,j,q}}} \right)}
\end{equation}
Denote  $Q$ as the number of subcarriers on the $jth$ RB. $B$ is the bandwidth of subcarrier.

First, we establish the optimal joint multi-cell RA model, aimed at maximizing the weighted throughput. It can be formulated as follows:

\begin{equation}
\begin{array}{l}
\arg \mathop {\max }\limits_{\forall {\rho _{_{{i_{{u}}}}}}} \sum\limits_{u = 1}^U {\sum\limits_{i = 1}^{{K_u}} {{w_{{i_{{u}}}}}\sum\limits_{j \in {\rho _{_{{i_{{u}}}}}}} {{R_{{i_{u}},j}}} } } \\
s.t.\\
\begin{array}{*{20}{c}}
{}&{{\rho _{{i_{{1}}}}} \cap {\rho _{{{i'}_{{1}}}}}}
\end{array} = \phi ,{i_{{1}}} \ne {{i'}_{{1}}}\\
\begin{array}{*{20}{c}}
{}&{{\rho _{{i_{{2}}}}} \cap {\rho _{{{i'}_{{2}}}}} = }
\end{array}\phi ,{i_{{2}}} \ne {{i'}_{{2}}}\\
\begin{array}{*{20}{c}}
{\begin{array}{*{20}{c}}
{\begin{array}{*{20}{c}}
{}&{}
\end{array}}&{\begin{array}{*{20}{c}}
{}&{}
\end{array}}
\end{array}}& \cdots
\end{array}\\
\begin{array}{*{20}{c}}
{}&{{\rho _{{i_{{U}}}}} \cap {\rho _{{{i'}_{{U}}}}} = }
\end{array}\phi ,{i_{{U}}} \ne {{i'}_{{U}}}
\end{array}
\end{equation}
the ${\rho_{{i_{u}}}}$ denotes the consecutive RBs assigned to the ${ith}$ user in the ${uth}$ cell. ${w_{{i_{u}}}}$ is the weighed
metric, denoting the fairness. In regard to Max SINR scheduling, it is a constant. Considering PF criterion, it can be written as ${w_{{i_{u}}}} = {1 \mathord{\left/ {\vphantom {1 {{{\overline r }_{{i_{u}}}}}}} \right. \kern-\nulldelimiterspace} {{{\overline r }_{{i_{u}}}}}}$. ${\overline r _{{i_{u}}}}$ is long-term average throughput for the ${ith}$ in the ${uth}$ cell.
The model constraints guarantee the consecutive RBs assigned to users are distinct, in order to make sure the subcarrier adjacency.

However, that model is with high complexity, and can only be solved by exhaustive search \cite{b11}. In next subsection, we will propose a pure BIP model, which can be solved by BBS instead of exhaustive search.

\subsection{Pure Binary-Integer Programming Model}

We define the concept of inter-cell UP at beginning, which is the $U$ users  $\left\{ {{i_1},{i_2}, \cdots ,{i_U}} \right\}$ in the cooperative scheduling network simultaneously transmit on the same RB, one user for each cell. In the single-cell LTE uplink RA, the individual user is the basic element to be scheduled. Nevertheless, considering the ICI mitigation, we should analyze the scheduling process of inter-cell UP, instead of individual user.  Moreover, we introduce the exclusivity constraint, indicating that all RBs will be allocated finally. Specially, in order to ensure both the exclusivity and adjacency, we adopt  the consecutive resource pattern as the basic RA element \cite{b11}. If the total RBs are $F$, the number of resource patterns is $J = {{\left( {{F^2} + F} \right)} \mathord{\left/ {\vphantom {{\left( {{F^2} + F} \right)} 2}} \right. \kern-\nulldelimiterspace} 2}$.

In this pure BIP model, the matrixes ${\bf{A}}$ and ${\bf{B}}$ are introduced to ensure adjacency and exclusivity. There are ${N_{all}} = {K_1} \times {K_2} \times  \cdots  \times {K_U}$ different ways to choose $U$ users to constitute the inter-cell UP. For integrative consideration of all users' CSI, all inter-cell UPs' combinations are enumerated. The reward vector ${\bf{C}} = {\left[ {{c_{1,1}}, \cdots ,c{}_{1,J}, \cdots ,{c_{{N_{all}},1}}, \cdots ,{c_{{N_{all}},J}}} \right]^T}$, and we will introduce the metric in detail in next subsection. The binary-integer variable vector containing $J \times {N_{all}}$ elements is correspondingly written, ${\bf{x}} = {\left[ {{x_{1,1}}, \cdots ,{x_{1,J}}, \cdots ,{x_{{N_{all}},1}}, \cdots ,{x_{{N_{all}},J}}} \right]^T}$.  We can formulate this pure BIP model as follows.
\begin{equation}
\begin{array}{l}
\begin{array}{*{20}{c}}
{\max }&{{{\bf{C}}^T}{\bf{x}}}
\end{array}\\
\begin{array}{*{20}{c}}
{s.t.}&\begin{array}{l}
{\bf{Ax}} = {\bf{1}}\\
{\bf{Bx}} \le {\bf{1}}\\
{x_{i,j}} \in \left\{ {0,1} \right\}
\end{array}
\end{array}
\end{array}
\end{equation}
Here, generally, all inter-cell UPs are in response to the same consecutive resource pattern, so the basic resource pattern matrix $\bf{T}$ is the same, that is ${\bf{A}} = {{{\bf{1}}_{{N_{all}}}}} \otimes {\bf{T}}$. The symbol $ \otimes $ is the Kronecker product operator.
Each column of basic resource pattern matrix $\bf{T}$ stands for one resource pattern, and the whole basic resource pattern matrix $\bf{T}$ exhaustively enumerates all the resource patterns. We can describe the ${lth}$ column of $\textbf{T}$ representing the ${lth}$ resource pattern as follows.

\begin{equation}
{{\bf{T}}_{:,l}} = {\left[ {{{\bf{0}}_{x - 1}},{{\bf{1}}_L},{{\bf{0}}_{F - L - x + 1}}} \right]^T}
\end{equation}
It includes the $L$ RBs. Specially, ${0_0} = \phi$ denotes empty vector. For the ${lth}$ column, the fist digit one begins at the ${xth}$ row.
\begin{equation}
x = \left\{ {\begin{array}{*{20}{c}}
l&{1 \le l \le F}\\
{l - F}&{F < l \le 2F - 1}\\
 \vdots & \vdots \\
{l - \sum\limits_{f = 1}^{F - 1} {\left( {F - \left( {f - 1} \right)} \right)} }&{\sum\limits_{f = 1}^{F - 1} {\left( {F - \left( {f - 1} \right)} \right)}  < l \le J}
\end{array}} \right.
\end{equation}
In addition, if the number of RB is 3, the matrix ${\bf{T}}$ can be:

\begin{center}
$
\textbf{T}=\left[ {\begin{array}{*{20}{c}}
1&0&0&1&0&1\\
0&1&0&1&1&1\\
0&0&1&0&1&1
\end{array}} \right]
$
\end{center}
${\left[ {1,0,0} \right]^{\rm{T}}}$ means the first RB is pitched on. ${\left[ {1,1,1} \right]^{\rm{T}}}$ denotes all RBs will be assigned. If the number of RB is 4, the matrix ${\bf{T}}$ will be:

\begin{center}
$
\textbf{T}=\left[ {\begin{array}{*{20}{c}}
1&0&0&0&1&0&0&1&0&1\\
0&1&0&0&1&1&0&1&1&1\\
0&0&1&0&0&1&1&1&1&1\\
0&0&0&1&0&0&1&0&1&1
\end{array}} \right]
$
\end{center}

Since inter-cell UP combines with resource pattern, so ${\bf{B}} = {\bf{D}} \otimes {{\bf{1}}_J}$. Moreover, ${\bf{D}}$ is the basic inter-cell UP matrix, indicating all combinations of inter-cell UP. For cooperative scheduling systems containing $U$ cells, it can be expressed as follows.
\begin{equation}
{\bf{D}} = \left[ {\begin{array}{*{20}{c}}
{{E_{K_1}} \otimes {{\bf{1}}_{{K_2} \times  \cdots  \times {K_U}}}}\\
\begin{array}{l}
{{\bf{1}}_{K1}} \otimes \left( {{E_{{K_2}}} \otimes {{\bf{1}}_{{K_3} \times  \cdots  \times {K_U}}}} \right)\\
\begin{array}{*{20}{c}}
{\begin{array}{*{20}{c}}
{}&{}
\end{array}}&{\begin{array}{*{20}{c}}
{}&\begin{array}{l}
 \cdot \\
 \cdot \\
 \cdot
\end{array}
\end{array}}&{\begin{array}{*{20}{c}}
{}&{\begin{array}{*{20}{c}}
{}&{}
\end{array}}
\end{array}}
\end{array}\\
{{\bf{1}}_{{K_1} \times  \cdots  \times {K_{U - 2}}}} \otimes \left( {{E_{{K_{U - 1}}}} \otimes {{\bf{1}}_{{K_U}}}} \right)
\end{array}\\
{{{\bf{1}}_{{K_1} \times  \cdots  \times {K_{U - 1}}}} \otimes {E_{{K_U}}}}
\end{array}} \right]
\end{equation}
${E_K}$ is the $K$ dimension identity matrix. The digit one represents that this user is picked up to constitute the inter-cell UP. Take three cells cooperation for example, if ${K_1} = 2, {K_2} = 4,{K_3} = 1$, the ${\bf{D}}$ is:

\begin{center}
$
\textbf{D}=\left[ {\begin{array}{*{20}{c}}
1&1&1&1&0&0&0&0\\
0&0&0&0&1&1&1&1\\
1&0&1&0&1&0&1&0\\
0&1&0&1&0&1&0&1\\
1&1&1&1&1&1&1&1
\end{array}} \right]
$
\end{center}
In addition, take two cells cooperation for example, if ${K_1} = 3, {K_2} = 3$, the ${\bf{D}}$ is formulated as follows:
\begin{center}
$
\textbf{D}=\left[ {\begin{array}{*{20}{c}}
1&1&1&0&0&0&0&0&0\\
0&0&0&1&1&1&0&0&0\\
0&0&0&0&0&0&1&1&1\\
1&0&0&1&0&0&1&0&0\\
0&1&0&0&1&0&0&1&0\\
0&0&1&0&0&1&0&0&1
\end{array}} \right]
$
\end{center}

What deserve attention is to ensure the subcarrier adjacency we add some restrictions. If one user has obtained one resource pattern, it can't scheduled once again in this transmission time interval (TTI). Whereas as long as that solution can guarantee subcarrier adjacency, it will be feasible. That characteristics can be summarized as inter-cell UPs alignment for resource, which is beneficial for lowering computational complexity but also decreases the performance. Since we conservatively remove some feasible solutions makes pure BIP model to be suboptimal.

\subsection{Metric for Throughput and Fairness}

We introduce the metric based on the PF criterion for throughput and fairness in this subsection, which is designed to let the inter-cell UP with higher instantaneous throughput and lower long-term average throughput be scheduled. The metric balances the throughput and fairness, and a fairness factor is given to be a tradeoff point. Besides, the metric can be seemed as the sum of single RB and user PFs. At last, the combinations of metric are with a linear relationship.

For the ${pth}$ inter-cell UP with  the ${lth}$ resource pattern, the metric can be written as follows.
\begin{equation}
{c_{p,l}} = \sum\limits_{u = 1}^U {{{\left( {\frac{1}{{{{\bar r}_{{i_u}}}}}} \right)}^\delta }}  \cdot \sum\limits_{j = x}^{x + L - 1} {{R_{{i_u},j}}}
\end{equation}
${{R_{{i_u},j}}}$ is the instantaneous throughput for $i_uth$ user at $jth$ RB. The fairness factor $\delta$ is designed to adjust the importance of fairness. ${\overline r _{{i_u}}}\left( t \right) $ is the long-term average throughput for the $i_uth$ user at TTI $t$.
\begin{equation}
{\overline r _{{i_u}}}\left( t \right) = \beta  \cdot {\overline r _{{i_u}}}\left( {t - 1} \right) + \left( {1 - \beta } \right) \cdot {R_{{i_u}}}\left( {t - 1} \right)
\end{equation}
In that formula, $\beta$ is the fairness forgetting factor, ${R_{{i_u}}}\left( t-1 \right)$ is the served throughput at TTI $t-1$. In fact, ${R_{{i_u}}}\left( t-1 \right) = 0$ when the $i_uth$ user isn't scheduled at TTI $t-1$.And that also will be Max SINR scheduling, if we only take instantaneous throughput into consideration. That is ${{{\overline r }_{{i_u}}}}$ becomes a constant.

\begin{table*}
\begin{center}
\caption{Average spectrum efficiency and 5th percentile of user throughput}
\begin{tabular}{c|c|c|c|c}
  \hline
  \rowcolor[gray]{0.8}[5.8pt][5.8pt]& Average spectrum efficiency & Gains & 5th percentile of user throughput & Gains \\
  \hline
  Single-cell optimal RA \cite{b11} & 2.9427 bps\textbackslash Hz\textbackslash sector & / & 1644.5057 kbps\textbackslash user& / \\
  \hline
  Multi-cell cooperative RA \cite{b6} & 3.2183 bps\textbackslash Hz\textbackslash sector & 9.37\% & 1794.9339 kbps\textbackslash user & 9.15\% \\
  \hline
  Proposed RA & 3.3790 bps\textbackslash Hz\textbackslash sector & 14.83\% & 2008.5080 kbps\textbackslash user & 22.13\% \\
  \hline
\end{tabular}
\end{center}
\end{table*}
\FloatBarrier

\section{Simulation Results}

\begin{table}
%\begin{center}
\caption{The system-level simulation parameters}
\begin{tabular}{l|l}
  \hline
  Parameters & Settings \\
  \hline
  Inter-BS distance & 500m\\
  Carrier frequency & 2G\\
  Bandwidth & 10MHz\\
  Channel model & 3GPP Spatial Channel Model \\
  Path loss (PL) & according to \cite{b14} Case 1\\
  User speed & 3km\\
  Thermal noise & -174dBm\\
  HARQ & 0\\
  Throughput calculation& EESM\\
  Traffic model & Infinite full buffer\\
  Channel estimation & Ideal\\
  Power control & ${P_{\max }} = 24dBm,{P_0} =  - 60dBm,\alpha  = 0.6$\\
  \hline
\end{tabular}
%\end{center}
\end{table}

In this section, we introduce the system-level simulation parameters and the open-loop PC. Then through the system-level LTE uplink simulation platform we established, the extensive performance of our proposed RA scheme, multi-cell cooperative RA \cite{b6} and single-cell optimal RA \cite{b11} under the open-loop PC are investigated.

For approaching practical system, we employ the scenario 1 cell site with 3 sectors per site. Each sector site and user equipment are equipped with 2 antennas and 1 antenna, respectively. That deployment scenario can be seen as three cells cooperative scheduling network depicted in Figure 1. The remainder of parameters are listed in table II.

\subsection{Open-loop Power Control}

In regard to PC, we set the transmit power of one user according to an open-loop PC equation defined by:
\begin{equation}
P = \min \left\{ {{P_{\max }},{P_0} + 10 \cdot {{\log }_{10}}M + \alpha  \cdot PL} \right\}
\end{equation}
$P_{\max }$ is maximum user transmission power. Here, it is set to 23 dBm for 3GPP case 1 \cite{b14}. ${P_0}$ is the cell specific power offset. $M$ is the number of assigned RBs to user. $\alpha$ is power compensation factor, which can be set within $\left\{ {0,0.4,0.5,0.6,0.7,0.8,0.9,1} \right\}$. $\alpha=0$ means no power compensation and all users transmission power is the same. $\alpha=1$ denotes full power compensation and that makes best of system ability to balance PL. In practical, PL can be estimated through the downlink. The open-loop PC is an inverse PC. Cell-edge user will transmit at a higher power per RB than cell-center user, which is aimed at combating near-far effect.

\subsection{Performance Comparison}
\begin{figure}
\centering
\includegraphics[width=3.5in]{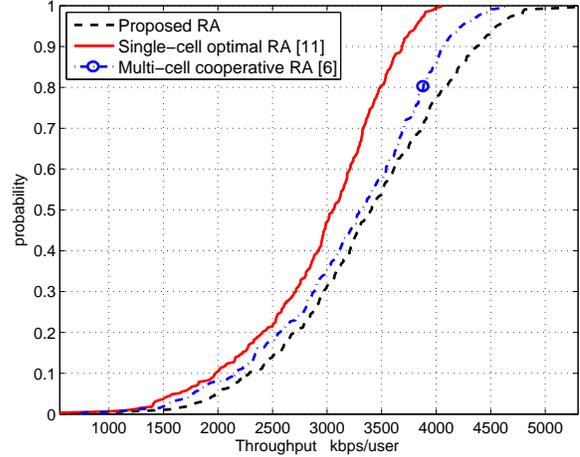}
\caption{The cumulative distribution functiuon (CDF) of user throughput.}
\end{figure}

\begin{figure}
\centering
\includegraphics[width=3.5in]{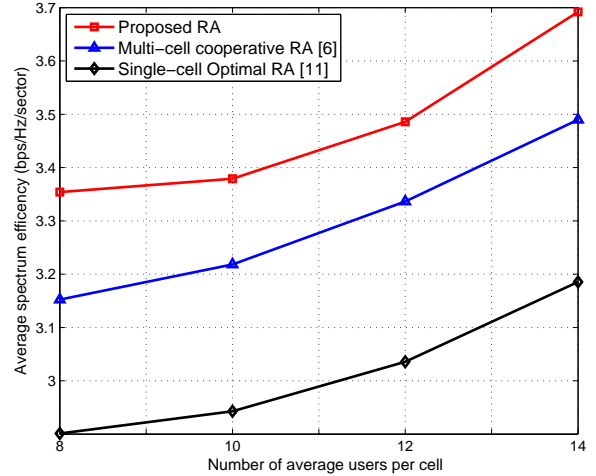}
\caption{Average spectrum efficiency with varying number of users.}
\end{figure}

We assume each cell is averagely assigned to 10 users, and compare the performance of our proposed RA, multi-cell cooperative RA \cite{b6} and single-cell optimal RA \cite{b11}. That simulation results are shown in Figure 2 and Table I. In Figure 2, due to the strong ICI, although the single-cell optimal RA enjoys a more reasonable RB allocation strategy under no interference assumption, herein the multi-cell cooperative RA still outperforms the single-cell optimal RA in all users, which illustrates that ICI degrades the performance badly in multi-cell scenes. Furthermore, the best performance of our proposed RA scheme in regard to cell-center user and cell-edge user is also shown. This is because our proposed RA  is with the integrative consideration of all CSI and better efficient RB allocation strategy. Figure 2 also depicts that cell-center users can also generate a large gain to other schemes, which owes to carefully RB allocation. The system fairness is represented by the cell-edge users (5th percentile of user user in CDF) in our paper. In Table I, in contrast to those comparison schemes, our proposed RA yields significant gains in 5th percentile of user throughput, because cell-edge users benefit a lot from ICI mitigation design. At last, we can find that the average spectrum efficiency is also improved prominently in our proposed RA scheme.

In Figure 3, with the increasing number of users in three cells cooperation scheduling network, the average spectrum efficiency of all RA schemes improve  simultaneously. The RA scheme is more frequently to choose the higher throughput user at large number of users, which can be summarized as multi-user diversity \cite{b15}. Furthermore the gains of proposed RA will be maintained, when the number of users is not 10 average users.

\section{conclusion}
We introduce a joint multi-cell RA using pure BIP scheme, and compare it with single-cell optimal RA scheme and a multi-cell cooperative RA scheme. The simulation results show that our proposed RA scheme outperforms those two comparison schemes and yields significant gains in  5th percentile of user throughput and average spectrum efficiency. Furthermore, if we take the joint reception (JR) into consideration, the joint three cells RA using pure BIP scheme is also suitable for intra-site uplink CoMP RA.

\section{Acknowledgement}
We want to thank all the project partners for the great collaboration within WTI laboratory. Furthermore, the work is supported by the National High Technology Research and Development Program of China (2014AA01A701)£¬ Beijing Nova program (No. xx2012037)  and Key National Science Foundation of China (61231009).


\begin{thebibliography}{99}
\bibitem{b1}
F. Jiancun, G. Y. Li, Y. Qinye, and L. Liangliang, \textquotedblleft Multiuser pairing and resource allocation with interference avoidance for SC-FDMA cellular systems, \textquotedblright \emph{in Proc. 2012 IEEE GLOBECOM.}
\bibitem{b2}
A. Ahmad and M. Assaad, \textquotedblleft Polynomial-Complexity Optimal Resource Allocation Framework for Uplink SC-FDMA Systems, \textquotedblright \emph{in Proc. 2011 IEEE GLOBECOM.}
\bibitem{b3}
X. Tao, F. Xu, W. Rehman, Y. Xu, and X. Li, \textquotedblleft A Generic Mathematical Model Based on Fuzzy Set Theory for Frequency Reuse in Cellular Networks, \textquotedblright \emph{ IEEE J. Sel. Areas Commun.}, vol. 31, no. 5, pp. 861-868, May 2013.
\bibitem{b4}
N. Himayat, S. Talwar, A. Rao,and  R. Soni, \textquotedblleft
Interference management for 4G cellular standards [WIMAX/LTE UPDATE],\textquotedblright \emph{IEEE Commun. Mag.}, vol. 48, pp. 86-92, 2010.
\bibitem{b5}
X. Zhang, Y. Li, and H. Ji. , \textquotedblleft An energy efficiency power and
sub-carrier allocation for the downlink multi-user CoMP in multi-cell systems, \textquotedblright \emph{in Proc. 2012 IEEE ICC.}
\bibitem{b6}
P. Frank, A. Muller, H. Droste,and J. Speidel, \textquotedblleft
Cooperative interference-aware joint scheduling for the 3GPP LTE uplink, \textquotedblright \emph{in Proc. 2010 IEEE PIMRC.}
\bibitem{b7}
3GPP TS 36.213 V11.4.0 Evolved universal terrestrial radio access (EUTRA): Physical layer procedure. Sept. 2013.
\bibitem{b8}
C. Xu, L. Song, Z. Han, Q. Zhao, X. Wang, and B. Jiao, \textquotedblleft Efficient Resource Allocation for Device-to-Device Underlaying Networks using Combinatorial  Auction, \textquotedblright \emph{ IEEE J. Sel. Areas Commun.},  vol. 31, no. 9, pp. 348-358, Sep. 2013.
\bibitem{b9}
M. Zhou, Q. Cui, R. Jantti, X. Tao, \textquotedblleft Energy-Efficient Relay Selection and Power Allocation for Two-Way Relay Channel with Analog Network Coding, \textquotedblright \emph{ IEEE Commun. Lett.} vol. 16, no. 6, pp. 816-819, Jun. 2012.
\bibitem{b10}
X. Tao, X. Xu, and Q. Cui, \textquotedblleft An overview of cooperative communications, \textquotedblright
\emph{IEEE Commun. Magazine,}  vo1.50, no.6, pp.65-71, June 2012.
\bibitem{b11}
I. C. Wong, O. Oteri, and W. McCoy,\textquotedblleft
Optimal resource allocation in uplink SC-FDMA systems,\textquotedblright \emph{ IEEE Trans. on Wireless Commun.}, vol. 8, pp. 2161-2165, 2009.
\bibitem{b12}
J. Fan, G. Y. Li, Y. Qinye, B. Peng, and Z. Xiaolong,\textquotedblleft
Joint User Pairing and Resource Allocation for LTE Uplink Transmission,\textquotedblright \emph{ IEEE Trans. on Wireless Commun.}, vol. 11, pp. 2838-2847, 2012.
\bibitem{b13}
3GPP TR 25.814 V7.1.0, \textquotedblleft Physcial layer aspects for evolved universal terrestrial radio access (UTRA),  \textquotedblright \emph{Sep. 2006.}
\bibitem{b14}
3GPP TR 36.814 V9.0.0, \textquotedblleft Further Advancements for E-UTRA Physical Layer Aspects, \textquotedblright \emph{Mar. 2010.}
\bibitem{b15}
R. Knopp and P. Humblet,  \textquotedblleft
Information capacity and power control in single-cell multiuser communications, \textquotedblright \emph{in Proc. 1995 IEEE ICC.}
\end{thebibliography}
\end{document}